\documentclass{article}

\usepackage{arxiv}

\usepackage{url}            
\usepackage{booktabs}       
\usepackage{amsfonts}       
\usepackage{nicefrac}       
\usepackage{microtype}      

 \usepackage{hyperref}
 \hypersetup{colorlinks=true,linkcolor=citeblue,citecolor=citeblue,urlcolor=citeblue}

\usepackage{cite}
\usepackage{amsmath,amssymb,amsfonts}
\usepackage{cleveref}
\usepackage{subcaption}
\usepackage{makecell}

\usepackage[T1]{fontenc}
\usepackage{lmodern}

\usepackage{lipsum}
\usepackage{xcolor}
\usepackage{graphicx}
\usepackage{geometry}
\usepackage{array}
\usepackage{multirow}
\usepackage{xcolor}
\definecolor{citeblue}{RGB}{0,0,255}

\graphicspath{ {./images/} }

\title{Exploring Explainable AI Techniques for Improved Interpretability in Lung and Colon Cancer Classification}

\author{
\\
\newline
\\
Mukaffi Bin Moin\textsuperscript{*},
Fatema Tuj Johora Faria,
Swarnajit Saha,
Busra Kamal Rafa,
\\
Mohammad Shafiul Alam
\\
\bigskip
\\
\\Ahsanullah University of Science and Technology, Dhaka, Bangladesh.
\\
\\
\\
\bigskip
*Corresponding author(s). E-mail(s): \texttt{\textcolor{blue}{mukaffi28@gmail.com}}\\
Contributing authors: \texttt{\textcolor{blue}{fatema.faria142@gmail.com}}; \texttt{\textcolor{blue}{swarnajitsaha68@gmail.com}}; \\\texttt{\textcolor{blue}
{brafa263.3@gmail.com}}; \texttt{\textcolor{blue}
{shafiul.cse@aust.edu}}; 
\\ 
}

\begin{document}

\maketitle
\abstract{Lung and colon cancer are serious worldwide health challenges that require early and precise identification to reduce mortality risks. However, diagnosis, which is mostly dependent on histopathologists' competence, presents difficulties and hazards when expertise is insufficient. While diagnostic methods like imaging and blood markers contribute to early detection, histopathology remains the gold standard, although time-consuming and vulnerable to inter-observer mistakes. Limited access to high-end technology further limits patients' ability to receive immediate medical care and diagnosis. Recent advances in deep learning have generated interest in its application to medical imaging analysis, specifically the use of histopathological images to diagnose lung and colon cancer. The goal of this investigation is to use and adapt existing pre-trained CNN-based models, such as Xception, DenseNet201, ResNet101, InceptionV3, DenseNet121, DenseNet169, ResNet152, and InceptionResNetV2, to enhance classification through better augmentation strategies. The results show tremendous progress, with all eight models reaching impressive accuracy ranging from 97\% to 99\%. Furthermore, attention visualization techniques such as GradCAM, GradCAM++, ScoreCAM, Faster Score-CAM, and LayerCAM, as well as Vanilla Saliency and SmoothGrad, are used to provide insights into the models' classification decisions, thereby improving interpretability and understanding of malignant and benign image classification.
Our research implementations are open to the public at:
\href{https://github.com/Mukaffi28/Explainable-AI-for-Lung-and-Colon-Cancer-Classification}{{https://github.com/Mukaffi28/Explainable-AI-for-Lung-and-Colon-Cancer-Classification}}}



\section{Introduction}\label{introlab}
A group of disorders referred to as cancer is characterised by an uncontrolled expansion and spread of abnormal cells within the body. These cells can infiltrate surrounding organs and tissues, leading to dangerous health consequences \cite{Intro1}.  In 2023\footnote{\href{https://www.paho.org/en/campaigns/world-cancer-day-2023-close-care-gap}{ {https://www.paho.org/en/campaigns/world-cancer-day-2023-close-care-gap}}}, approximately 10 million people worldwide died from cancer-related causes, highlighting the ongoing global impact of this disease on public health. It is projected that the cancer death rate would increase to 60\% by 2035.\cite{IntroC3}.
 Histopathological imaging is critical for diagnosing and curing lung and colon cancers because it provides microscopic images of tissue samples, highlights cellular architecture, and informs treatment decisions \cite{IntroC3}. Digital pathology, powered by scanning technology, is transforming medicine by digitizing histopathological slides to improve disease diagnosis and management. AI integration increases its potential for clinical diagnosis and research \cite{Intro4}. This powerful combination has the potential to expand its applications in various fields, including cancer detection \cite{IntroC1}, cardiovascular diseases \cite{cardiovascular}, neurological disorders 
 \cite{neurological}, diabetic retinopathy \cite{diabetic}, pulmonary diseases \cite{pulmonary}, and skin diseases \cite{skin}.

In the last few years, major improvements have been achieved in the automated classification of lung and colon cancer using histopathological images. Through the use of convolutional neural networks (CNNs), Sanidhya's \cite{IntroC1} study demonstrates the promise of artificial intelligence (AI) in enhancing cancer diagnoses by developing a computer-aided diagnosis system that can accurately detect lung and colon tumors from histopathological pictures with high accuracy (for the lung, 97\%, and for the colon, 96\%.). Osamu et al. used convolutional and recurrent neural networks to classify gastric and colonic epithelial tumors in histopathology images. With AUCs of 0.96 \& 0.99 for colon cancer \& adenoma, and 0.97 and 0.99 to stomach cancer and adenoma, respectively, the models were incredibly accurate\cite{IntroC7}. In another study Sudhakar's \cite{Intro4} introduces an automated method using EfficientNetV2 models to detect lung and colon cancer subtypes with 99.97\% accuracy, surpassing current methods. Their approach, which includes gradCAM-generated visual maps, assists pathologists in identifying critical locations for treatment plans, and it shows promise for clinical automation in cancer detection. Neha's research \cite{IntroC2} uses CNNs (ResNet50, VGG19, InceptionResNetV2, DenseNet) to classify lung cancer histology, aiming to improve diagnosis accuracy for better treatment decisions and reduce pathologists' workload, potentially improving patient outcomes in lung cancer care. 

After thorough analysis, a noticeable gap emerges in the realm of explainable AI techniques applied to lung and colon cancer using histopathological images. In this research paper, we focus to automate the detection of lung and colon cancer using histopathological images. We undertake a thorough evaluation of eight renowned pre-trained CNN models, including Xception, DenseNet121, DenseNet169, DenseNet201, InceptionV3, ResNet101, ResNet152, and InceptionResNetV2. To improve interpretability, we use Explainable AI approaches as Grad-CAM, Grad-CAM++, Score-CAM, FasterScore-CAM, Layer CAM for Class Activation Maps, and Vanilla Saliency and SmoothGrad for Saliency Maps. These strategies clarify the logic behind our models' predictions, leading to a better understanding of their decision-making processes.
\begin{itemize}
    \item The inspection of eight pre-trained CNN models (Xception, DenseNet121, Dense\-Net169, DenseNet201, InceptionV3, ResNet101, ResNet152, and InceptionResNetV2) for automated lung and colon cancer classification in histological images. 
    \item Explainable AI approaches including Grad-CAM, Grad-CAM++, Score-CAM, FasterScore-CAM, Layer CAM, Vanilla Saliency, and SmoothGrad have been integrated to increase interpretability.
 \item  Our research attempts to close the gap in explainable AI methods for colon and lung cancer classification in histopathology images.
\end{itemize}

\section{Related Works} \label{Relatedlab}
\subsection{Lung and Colon Cancer Classification without using XAI Techniques}
Sanidhya et al. \cite{IntroC1} used CNNs to identify malignancies of the lung and colon, with diagnosis accuracy of more than 97\% for cancer of the lungs and 96\% for cancer of the colon, based on digital pathology images from the LC25000 dataset. Highlighting the vital necessity of accurate and timely lung cancer histology detection, Neha et al. \cite{IntroC2} builds upon prior work in lung cancer diagnosis and classification.  Deep learning techniques, particularly CNNs like ResNet 50, VGG-19, Inception\_ResNet\_V2, and DenseNet, have shown promise in analyzing histopathological images for accurate subtype classification. Hasan et al. \cite{IntroC3} offered a deep convolutional neural network model for precise identification of colon adenocarcinoma from digital histopathology images, achieving up to 99.80\% accuracy. Notably, none of these studies explored Explainable AI techniques in the context of lung and colon cancer diagnosis.

\subsection{Lung and Colon Cancer Classification using XAI Techniques}
Satvik and Somya \cite{Intro1} developed a system that automates lung as well as colon cancer identification using deep neural networks on images from histopathology. Employing eight pre-trained CNN models, they achieved remarkable accuracies ranging from 96\% to 100\%. For explainable AI, they utilized GradCAM and SmoothGrad techniques. Introducing a novel Bilinear-CNN-based model, another research effort \cite{Intro2} focuses on whole-slide images (WSIs) automated tissue segmentation  of lung cancer. This model addresses challenges posed by tumor heterogeneity. For explainable AI, GradCAM was utilized. Sudhakar et al. \cite{Intro4} proposed an automated method utilizing EfficientNetV2 models for detecting lung and colon cancer subtypes from histopathology images. Visual saliency maps were employed to aid in understanding model decisions, obtaining a remarkable 99.98\% maximum test accuracy on the LC25000 dataset. However, Ahmed et al. \cite{seven} offered a lightweight, CNN-based deep learning technique for accurate colon cancer detection. Despite lacking Class Activation Maps or Saliency Maps, their method achieved a high accuracy of 99.50\%, outperforming existing deep learning approaches.

\section{Dataset Description}
\begin{figure}[h]
    \centering
    \includegraphics[width=1.0\textwidth]{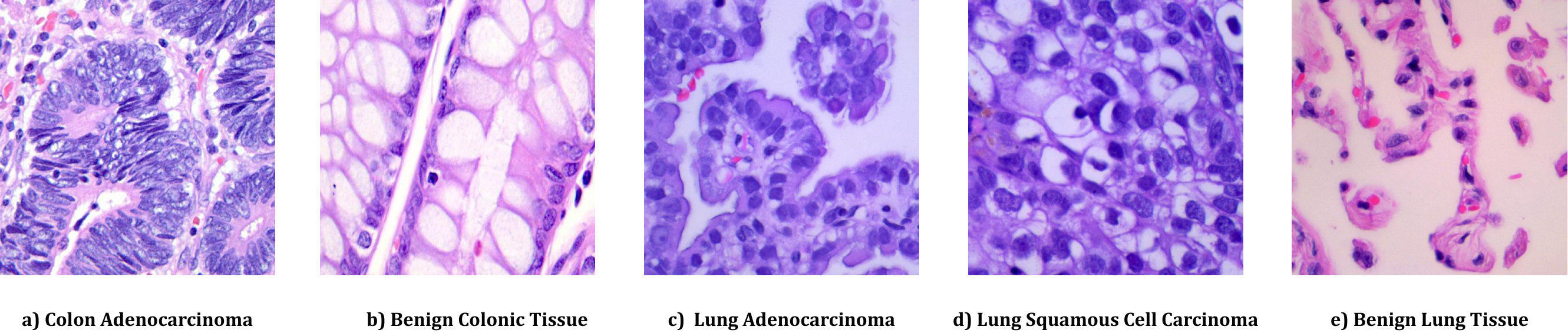}
    \caption{Representation of LC25000 dataset}
    \label{fig:data}
\end{figure}

We utilized the LC25000 dataset \cite{Dataset}, which carries 25,000 color pictures of lung and colon tissues categorised into five groups: lung squamous cell carcinoma, lung adenocarcinoma, benign lung tissue, colon cancer, and benign colonic tissue. 5,000 photos each class, each resized to 768 by 768 pixels. The collection is divided into colon and lung image sets in accordance with HIPAA compliance guidelines. It's instrumental in developing diagnostic tools for lung and colon cancers, driving progress in medical imaging research. The visual representation of the LC25000 dataset is displayed in Figure \ref{fig:data}.

\section{Background Study} \label{Backgroundlab}
\subsection{Convolutional Neural Networks (CNNs)}
Xception \cite{Xception}, DenseNet201 \cite{Dense}, DenseNet121 \cite{Dense}, DenseNet169 \cite{Dense}, ResNet101 \cite{Resnet}, ResNet152 \cite{Resnet}, InceptionV3 \cite{InceptionV3}, and InceptionResNetV2 \cite{inresV2} are all prominent Convolutional Neural Network (CNN) architectures. Xception, an ``Extreme Inception,'' enhances the Inception model by utilizing depthwise separable convolutions for improved computational efficiency. DenseNet models, including DenseNet201, DenseNet121, and DenseNet169, feature densely connected layers where each layer connects to every other layer, facilitating feature reuse. ResNet101 and ResNet152, part of the ResNet family, address the disappearing gradient issue by introducing skip connections, enabling the training of extremely deep networks. InceptionV3, a member of the Inception family, employs various filter sizes to capture features at multiple scales efficiently. InceptionResNetV2 combines the strengths of both Inception and ResNet architectures, integrating residual connections and multi-scale feature extraction for enhanced performance.

\subsection{Explainable Artificial Intelligence (XAI) techniques}
XAI techniques are essential for enhancing the transparency and interpretability of AI models. Among the notable methods are GradCAM, which identifies critical image regions by analyzing gradients of the target class score relative to convolutional neural network feature maps, and its extension GradCAM++, which refines localization accuracy by considering both positive and negative influences. ScoreCAM assigns importance scores to spatial locations in feature maps based on class scores, facilitating precise localization, while Faster Score-CAM optimizes efficiency for real-time applications. LayerCAM attributes importance scores to input pixels via relevance propagation across model layers, providing insights into decision-making processes. Gradients of output class scores are computed applying Vanilla Saliency in relation to input pictures, highlighting influential regions, whereas SmoothGrad enhances interpretability by averaging multiple saliency maps to reduce noise and provide smoother visualizations. 

\subsection{Evaluation Metrics}
In the evaluation of lung and colon classification tasks, several metrics are vital for comprehensively assessing model performance. Accuracy measures the overall correctness of predictions irrespective of class distribution, while precision and recall evaluate the model's ability to correctly identify instances of lung or colon conditions, focusing on minimizing false positives and false negatives, respectively. The F1 score strikes a balance between precision and recall, crucial for tasks with class imbalance. The Jaccard score assesses the overlap between predicted and actual classes, especially valuable in multi-class problems or imbalanced datasets. Finally, log loss quantifies the accuracy of predicted probabilities compared to actual probabilities, incentivizing well-calibrated predictions in lung and colon classification models.

\section{Proposed Methodology}
\begin{figure}
    \centering
    \includegraphics[width=1\textwidth, height=200px]{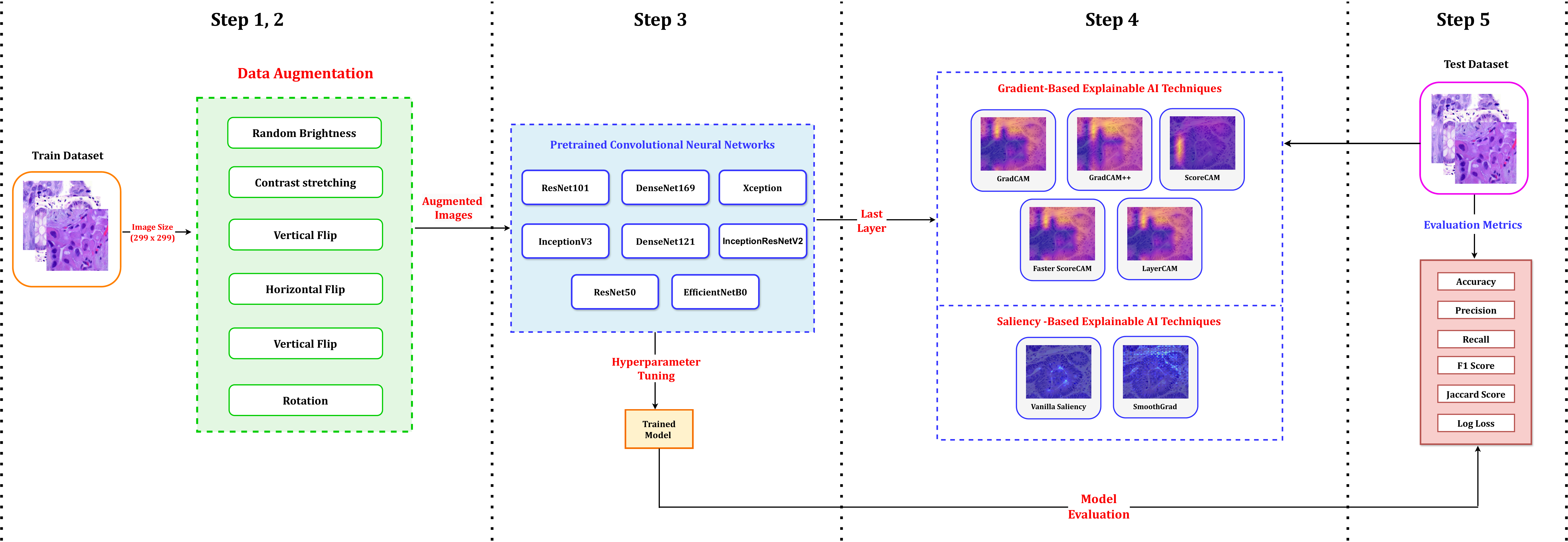}
    \caption{Representation of our proposed methodology for lung and colon cancer classification using explainable AI techniques}
    \label{fig:dia}
\end{figure}

This comprehensive approach allows us to compare the performance of various models in lung and colon cancer image classification tasks efficiently. Figure \ref{fig:dia} illustrates the methodology employed in our research.\\
\textbf{Step 1) Input Image:}
To ensure uniformity and compatibility across diverse models and techniques, the images underwent meticulous preprocessing. This crucial step involved resizing the images to a standardized dimension of 299x299 pixels.\\ 
\textbf{Step 2) Image Preprocessing:}
To prepare our dataset for model training, we initiated with pixel value normalization to a standard scale of 0 to 1, fostering convergence during training and addressing issues related to disparate data distributions. Following this, we introduced random rotations, diversifying the dataset and enhancing the model's adaptability to varied orientations. Horizontal flipping was also used to increase variety and reduce overfitting concerns by simulating mirror images. Subsequently, we selectively cropped images, eliminating extraneous background elements and emphasizing regions of interest, thereby facilitating feature extraction and reducing computational complexity. Addressing illumination discrepancies, we adjusted image brightness levels to ensure dataset consistency and minimize lighting variations' impact on model performance. Finally, we used contrast enhancement techniques to refine image contrast, particularly important in medical imaging where detailed information is critical for accurate analysis.\\
\textbf{Step 3) Model Selection and Training:}
For model selection, we opted for eight CNN architectures renowned for their efficacy in image classification tasks. These architectures include Xception, DenseNet201, ResNet101, InceptionV3, DenseNet121, DenseNet169, ResNet152, and InceptionResNetV2. Subsequently, we proceeded with model training by implementing each CNN architecture using the TensorFlow deep learning framework. The models were trained using the preprocessed dataset, which ensured coherence and stability throughout the training process. Furthermore, hyperparameter tuning was conducted, the details of which are provided in Table \ref{Hyperparameter}. This optimization process aimed to fine-tune the model's parameters and enhance its performance on the specific task of lung and colon cancer classification.
\\
\textbf{Step 4) Explainable AI Techniques:}
We applied eXplainable AI (XAI) techniques to the last layer of the CNN models to enhance interpretability and provide insights into model decision-making. For Class Activation Maps (CAM), we utilized GradCAM, GradCAM++, ScoreCAM, Faster Score-CAM, and LayerCAM to generate heatmaps highlighting discriminative regions in the images for each class, refining feature localization, producing class-specific attention maps, and visualizing activations at different network layers for deeper insights. Additionally, we employed Vanilla Saliency and SmoothGrad techniques for generating saliency maps to identify high-gradient regions influencing model predictions and enhancing interpretability.
\\
\textbf{Step 5) Performance Evaluation:}
Table \ref{lung_result} shows evaluation of each model to assess its effectiveness in lung and colon cancer classification. By employing these metrics, we gained a comprehensive understanding of each model's classification performance, allowing us to compare and select the most suitable model for the task at hand.

\section{Results and Discussion }\label{resultlab}
\subsection{Hyperparameter Settings}
Eight pretrained CNN models were trained to classify lung and colon cancer. To ensure optimal convergence without overfitting, an Adam optimizer with a batch size of ten and a learning rate that varied between 0.0001 and 0.001 throughout epochs 25 to 40 was employed by every model. Hyperparameter Settings for Pre-trained CNN Models for lung and colon cancer classification are shown in Table \ref{Hyperparameter}
\begin{table}[h]
\caption{Hyperparameter Settings for Pre-trained CNN Models for lung and colon cancer classification}
\vspace{5pt}
\label{Hyperparameter}
\centering
\begin{tabular}{p{2.6cm} p{2.0cm} p{1.8cm} p{2.5cm} p{1.8cm} } 
\toprule
\textbf{Models} & \textbf{Learning Rate} & \textbf{Batch Size} & \textbf{Number of Epochs} & \textbf{Optimizer} \\
\midrule
Xception & \centering 0.001 & \centering  10 & \centering 30 &  Adam \\
DenseNet201 & \centering 0.001 & \centering 10 & \centering 25 & Adam \\
ResNet101 & \centering 0.001 & \centering 10 & \centering 35 & Adam \\
InceptionV3 & \centering 0.001 & \centering 10 & \centering 25 & Adam \\
DenseNet121 & \centering 0.001 & \centering 10 & \centering 30 & Adam \\
DenseNet169 & \centering 0.001 & \centering 10 & \centering 35 & Adam \\
ResNet152 & \centering 0.001 & \centering 10 & \centering 35 & Adam \\
InceptionResNetV2 & \centering 0.001 & \centering 10 & \centering 40 & Adam \\
\bottomrule
\end{tabular}
\end{table}
\subsection{ Experiments}

\begin{figure}[h]
    \centering
  \begin{subfigure}[b]{0.23\textwidth}
    \includegraphics[width=\textwidth]{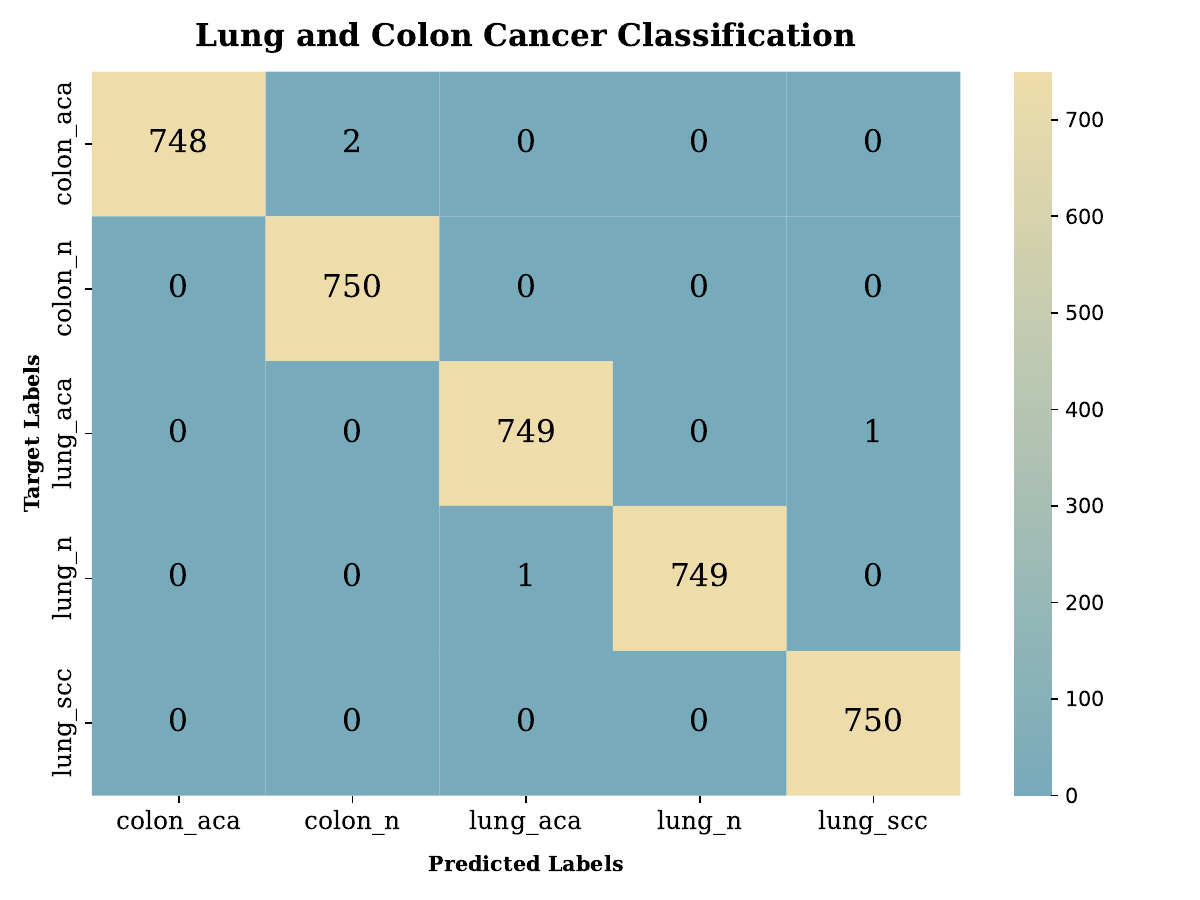}
    \caption{Xception}
  
  \end{subfigure}
  ~
  \begin{subfigure}[b]{0.23\textwidth}
    \includegraphics[width=\textwidth]{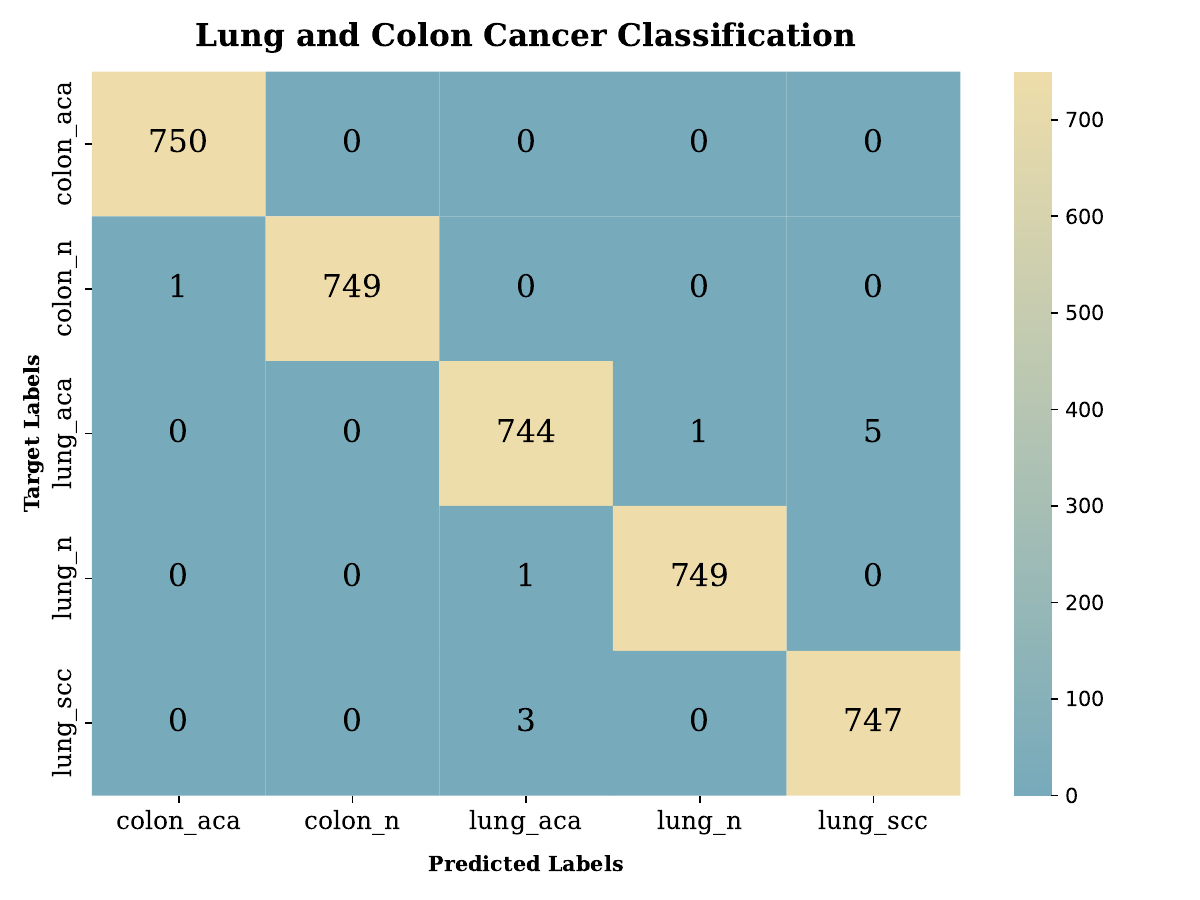}
    \caption{DenseNet201}
   
  \end{subfigure}
    ~
  \begin{subfigure}[b]{0.23\textwidth}
    \includegraphics[width=\textwidth]{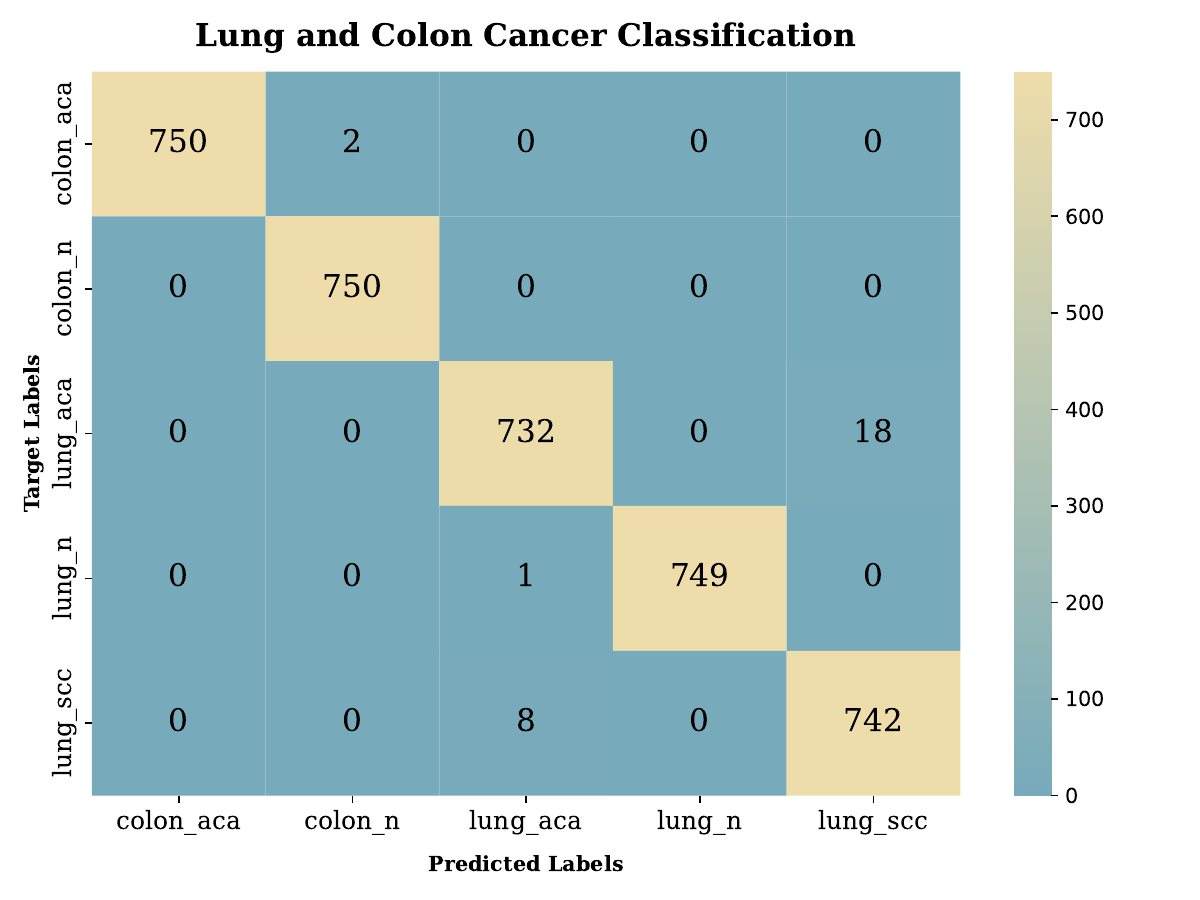}
    \caption{ResNet101}

  \end{subfigure}
    ~
  \begin{subfigure}[b]{0.23\textwidth}
    \includegraphics[width=\textwidth]{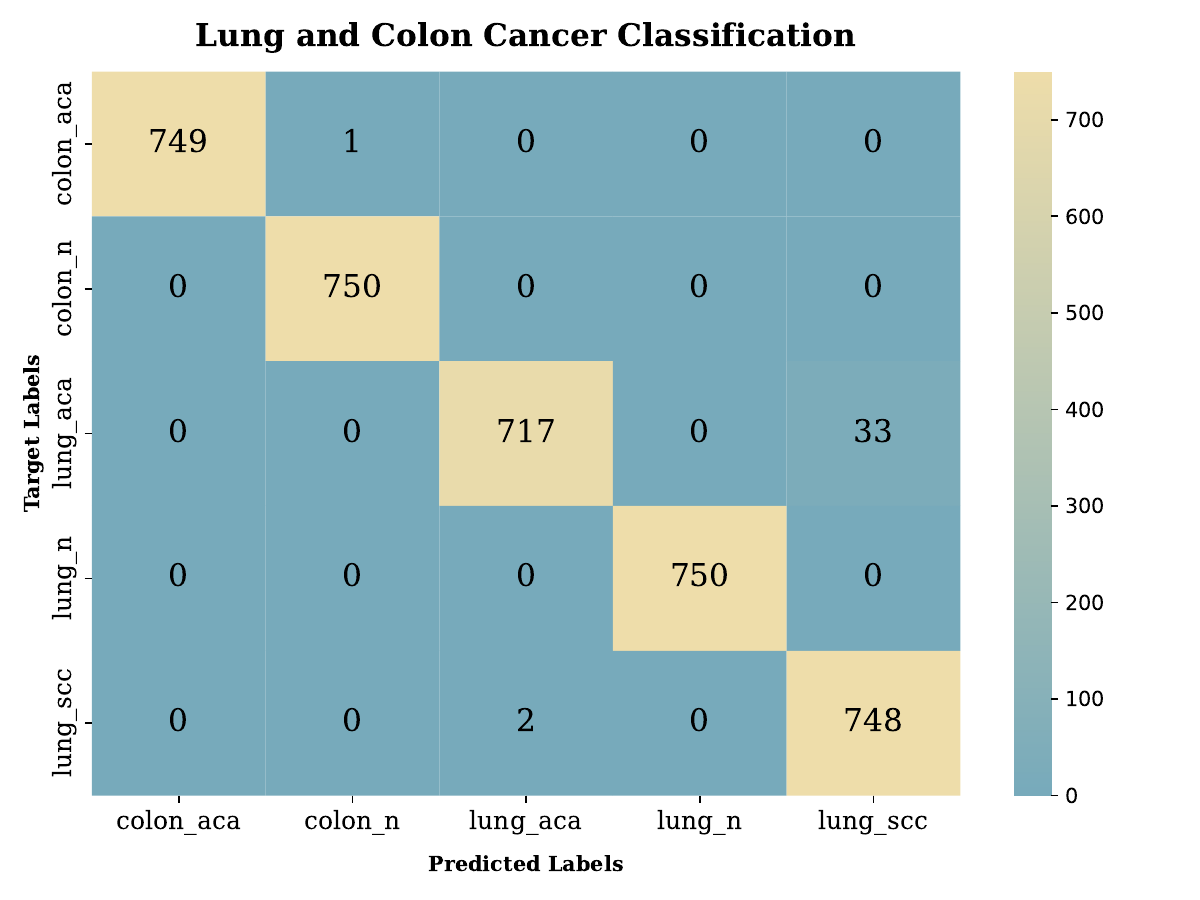}
    \caption{InceptionV3}

  \end{subfigure}

    \begin{subfigure}[b]{0.23\textwidth}
    \includegraphics[width=\textwidth]{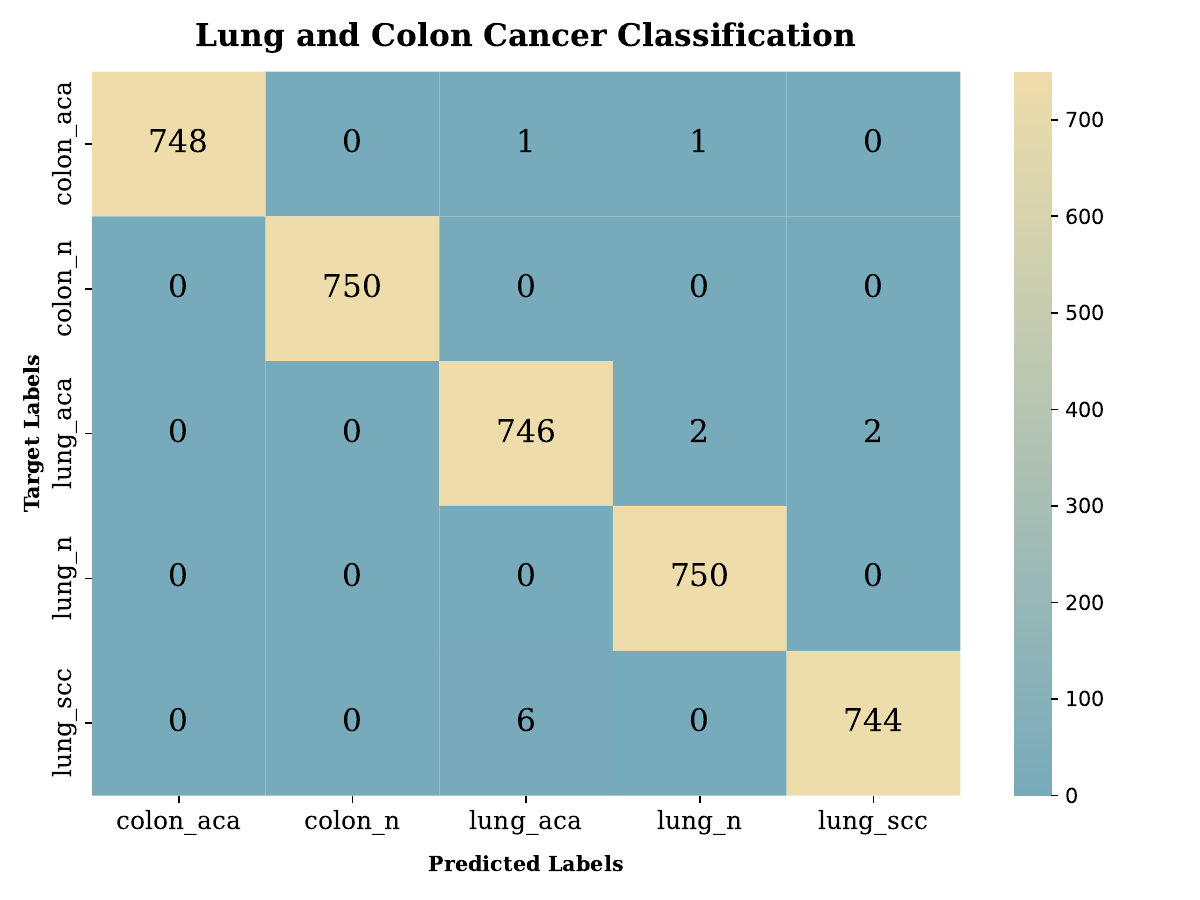}
    \caption{DenseNet121}

  \end{subfigure}
  ~
  \begin{subfigure}[b]{0.23\textwidth}
    \includegraphics[width=\textwidth]{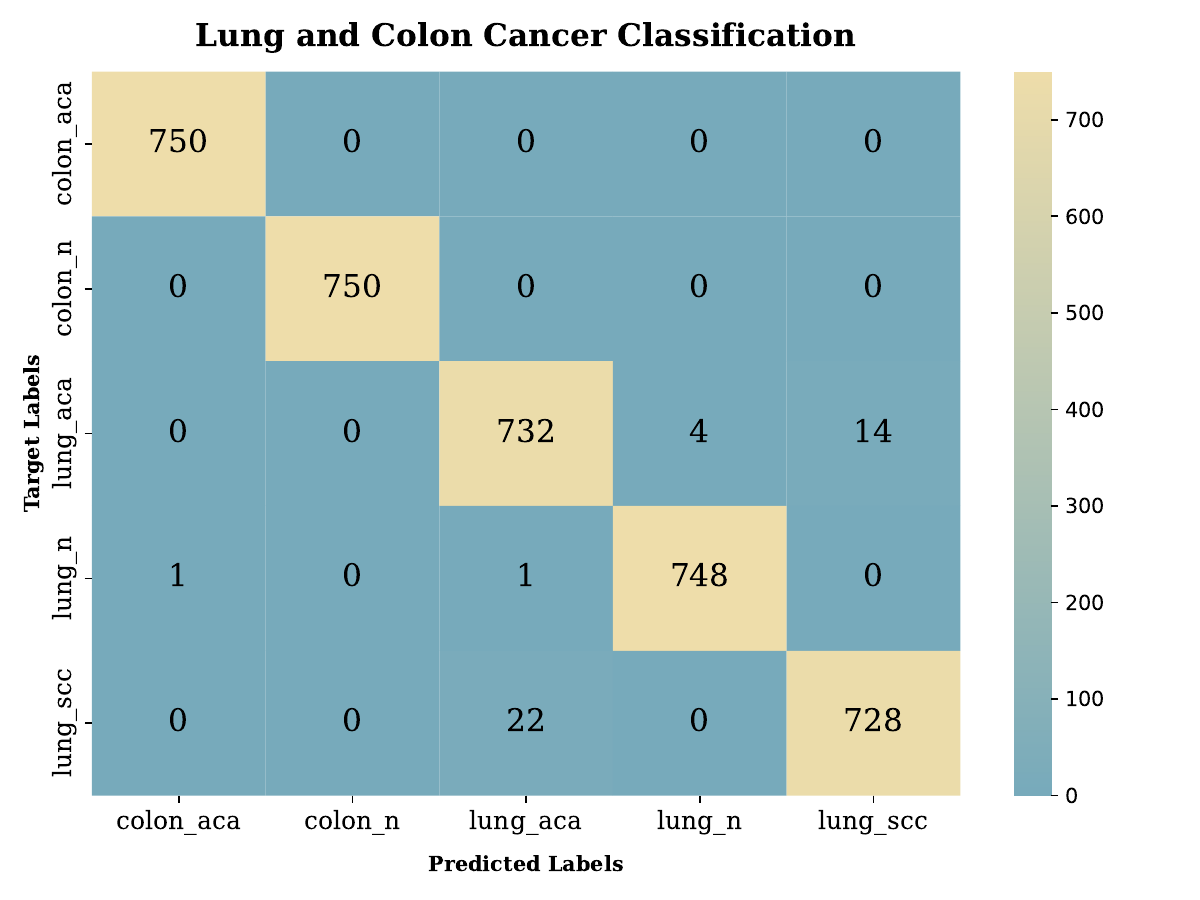}
    \caption{DenseNet169}

  \end{subfigure}
    ~
  \begin{subfigure}[b]{0.23\textwidth}
    \includegraphics[width=\textwidth]{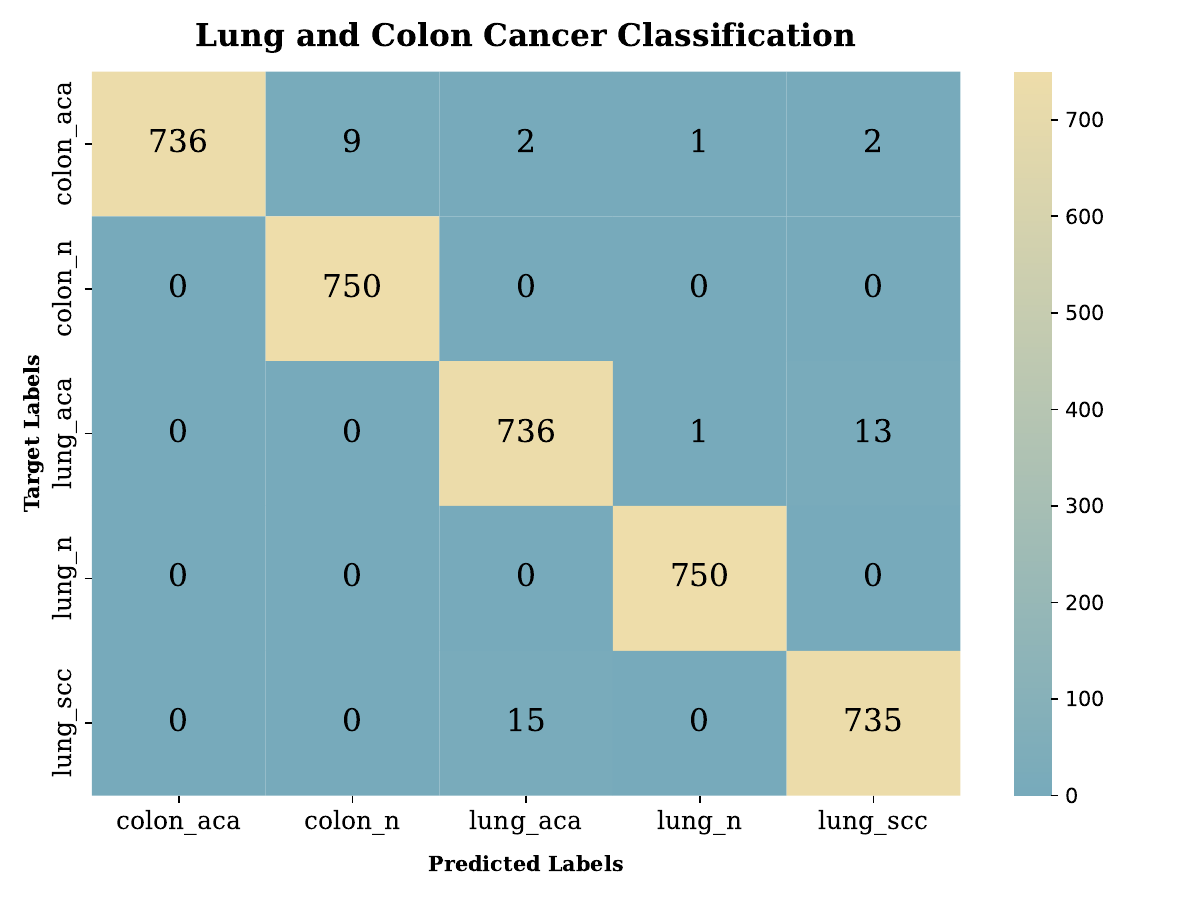}
    \caption{ResNet152}

  \end{subfigure}
    ~
  \begin{subfigure}[b]{0.23\textwidth}
    \includegraphics[width=\textwidth]{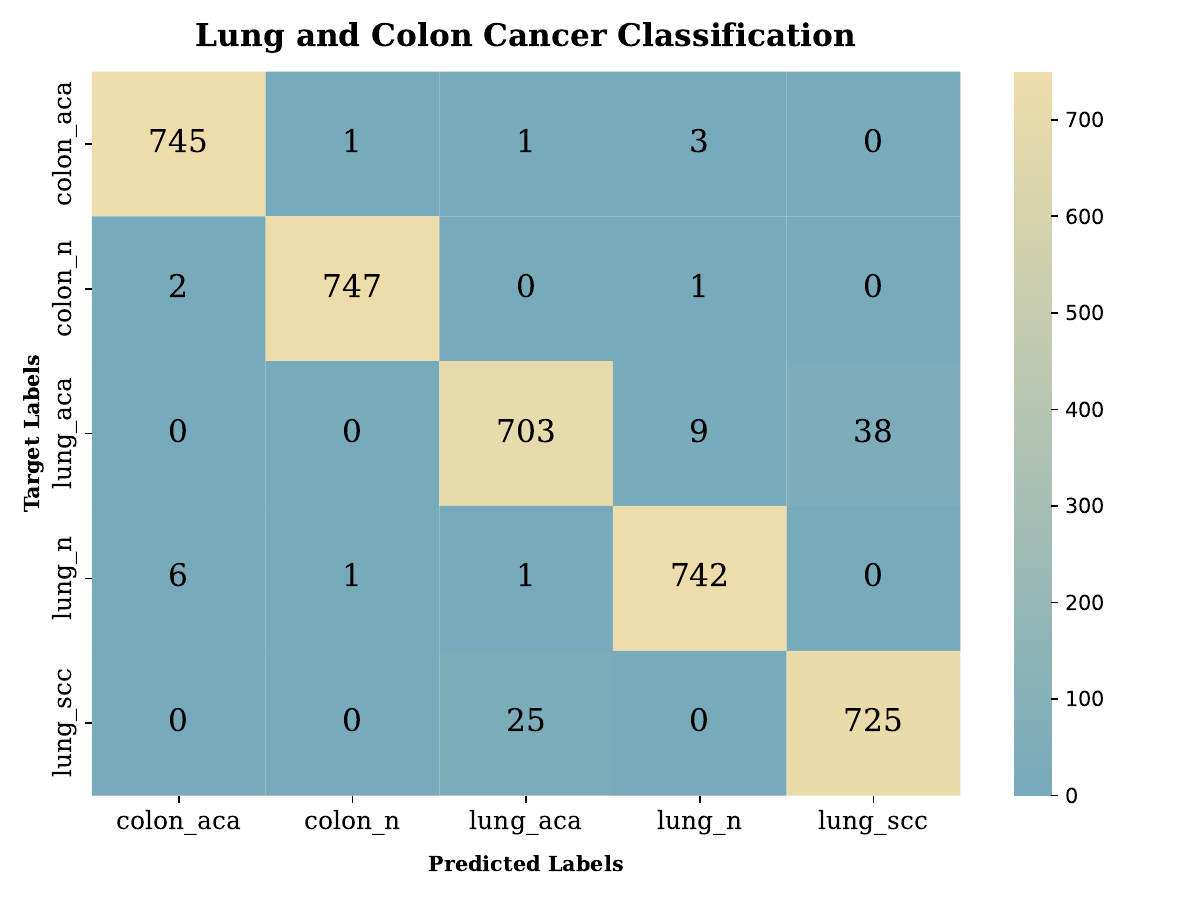}
    \caption{InceptionResNetV2}

  \end{subfigure}
    \caption{Confusion Matrices of Pretrained CNNs for Lung and Colon Cancer Classification}
    \label{fig:confusion_matrix}
\end{figure}
\begin{figure}[h]
    \centering
    \includegraphics[width=1.0\textwidth, height=190px ]{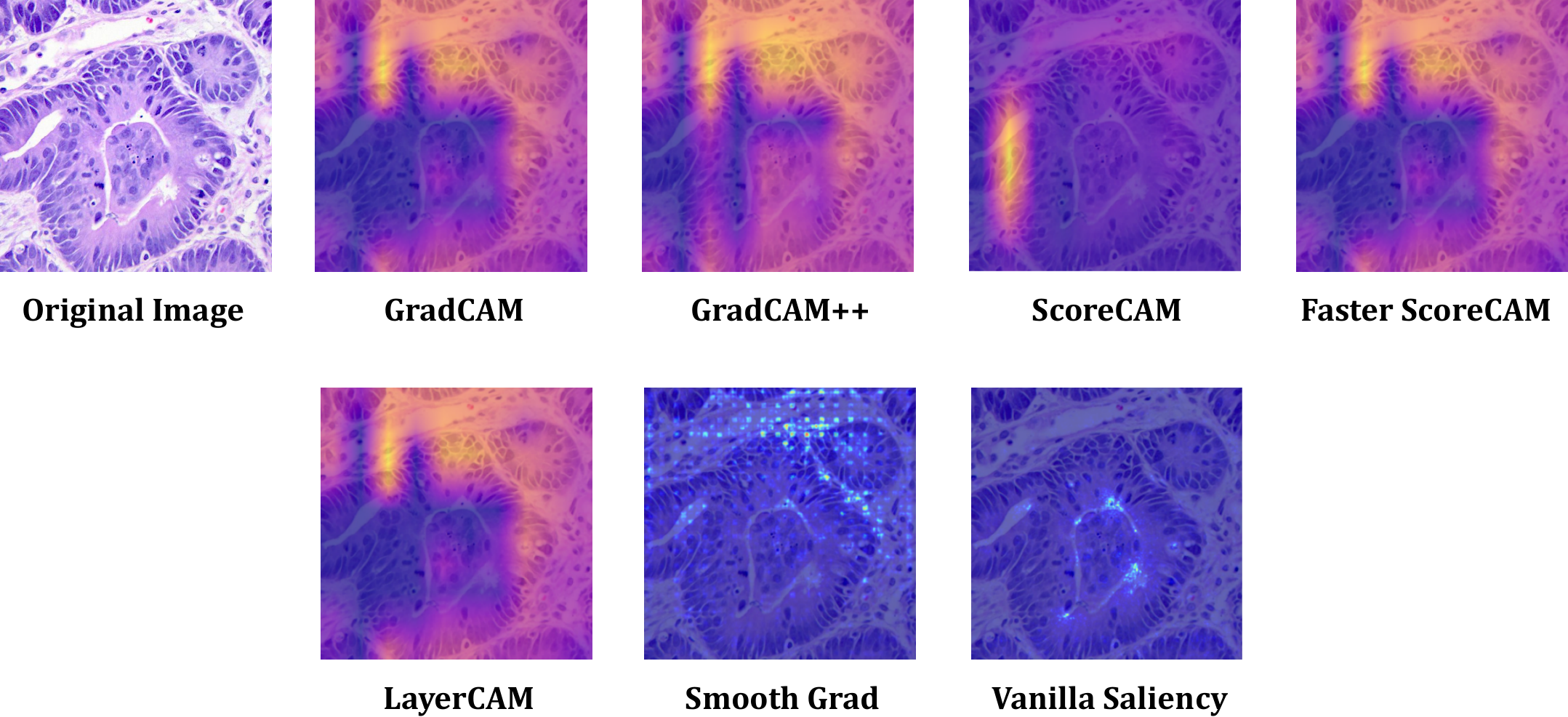}
    \caption{Different versions of AI insights such as GradCAM, GradCAM++, ScoreCAM, Faster ScoreCAM, LayerCAM, Smooth Grad, and Vanilla Saliency provide various explainable views of the original image in the context of lung and colon cancer classification}
    \label{fig:xai}
\end{figure}

The Table \ref{lung_result} compares performance for a total of eight pretrained CNN models in lung and colon cancer classification. With an accuracy of 0.9989 and a log loss of 0.0384, Xception performs the best.
 With the highest log loss of 0.8458 and the lowest accuracy of 0.9765, InceptionResNetV2 performs the worst. The metrics include accuracy, precision, recall, F1-score, Jaccard score, and log loss, with greater values signifying better outcomes except for log loss, where lower values are better. Figure \ref{fig:confusion_matrix} show Confusion Matrices of Pretrained CNNs for Lung and Colon Cancer Classification. Visualizations of the Various Explainable AI techniques are shown in Figure \ref{fig:xai}.

\begin{table}[h]
\caption{Performance Metrics Comparison of Various Pre-trained CNN Models}\label{lung_result}
\vspace{5pt}
\centering
\begin{tabular}{p{2.6cm} p{1.5cm} p{1.5cm} p{1.5cm} p{1.5cm} p{1.5cm} p{1.5cm} p{1.5cm} } 
\toprule
\textbf{Model} & \textbf{Accuracy} & \textbf{Precision} & \textbf{Recall} & \textbf{F1-Score} & \textbf{Jaccard Score} & \textbf{Log Loss} \\
\midrule
Xception  & 0.9989 & 0.9989 & 0.9989 & 0.9989 & 0.9978 & 0.0384\\
DenseNet201  & 0.9971&0.9971 & 0.9971&0.9971 & 0.9942 & 0.1057\\
ResNet101 & 0.9928&0.9928 &0.9928 &0.9928 & 0.9858 & 0.2595\\
InceptionV3  &0.9904 &0.9907 & 0.9904 &0.9904 & 0.9812 & 0.3460\\
DenseNet121 & 0.9896 &0.9898 & 0.9896& 0.9896 & 0.9795 &0.3749 \\
DenseNet169  &0.9888 &0.9888 &0.9888 &0.9888 &0.9781  &0.4037\\
ResNet152 & 0.9885 & 0.9886 & 0.9885 & 0.9885 & 0.9774 & 0.4133\\
InceptionResNetV2 & 0.9765 & 0.9765 & 0.9765 & 0.9765 & 0.9547  & 0.8458\\
\bottomrule
\end{tabular}
\end{table}


\section{Limitations and Future Research Directions}
\label{sec:6}
While our research has offered useful insights into the classification of lung and colon cancers, we note numerous limitations that must be considered. To begin, characteristics such as dataset size, quality, and variety can all have an impact on our models' effectiveness. To address these constraints, bigger and more diversified datasets would be required, maybe combining data from different sources to improve model generalization. Furthermore, our research concentrated on image-based categorization, ignoring possible synergies with other modalities such as genomes or clinical data. Future research might investigate multimodal techniques to increase classification accuracy and adaptability. Moreover, the interpretability of our models, while enhanced through XAI techniques, remains a challenge, particularly in complex medical domains. Further research is needed to develop more interpretable models and refine existing XAI methods to provide deeper insights into model decision-making processes. Furthermore, our research largely focused on two forms of cancer: lung and colon. Extending our method to additional forms of cancer would increase its relevance and influence in the area of oncology. Furthermore, examining our models' transferability to various healthcare settings and patient demographics will be beneficial for real-world deployment.

\section{Conclusion}
\label{sec:7}
Our research demonstrates the efficacy of our suggested methodology for lung and colon cancer classification, which uses advanced deep learning and eXplainable AI (XAI) techniques to improve accuracy and interpretability. We obtained outstanding classification results by applying cutting-edge CNN architectures to a standardized dataset. Notably, Xception outperformed all other models tested, with an amazing accuracy of 0.9989 and the lowest Log Loss of 0.0384. In contrast, InceptionResNetV2 had slightly lower accuracy (0.9765) and the highest Log Loss (0.8458), indicating that its classification capabilities should be improved. Our incorporation of XAI techniques such as GradCAM, GradCAM++, ScoreCAM, Faster Score-CAM, LayerCAM, Vanilla Saliency, and SmoothGrad yielded useful insights into the CNN models' decision-making processes. These strategies improved the visualization of discriminative areas in histopathological images, allowing for the detection of critical characteristics that contribute to the categorization of malignant and benign tissues. Such visualizations improve interpretability and enable more informed decision-making in healthcare settings, resulting in better patient care and treatment results. Overall, our study illustrates the power of sophisticated deep learning models and XAI approaches in medical image analysis, providing a solid foundation for precise cancer classification. Xception's superior performance demonstrates its efficacy in making confident and precise predictions, while insights from XAI techniques improve the interpretability of the classification process, establishing the way for improved diagnostic accuracy and clinical decision support in cancer treatment.

\bibliographystyle{unsrt}
\bibliography{Mukaffi}

\end{document}